\documentstyle[11pt]{article}
\begin{document}

\baselineskip=15pt
\rightline{UMN-TH-1249-94}
\rightline{hep-ph/yymmddd}
\rightline{April 1994}
\vskip .2in
\begin{center}
{\large{\bf Big Bang Baryogenesis\footnote{To be published in the
proceedings of the 33rd International Winter School on Nuclear and Particle
Physics, ``Matter Under Extreme Conditions", Feb. 27 - March 5 1994, Schladming
Austria, eds. H. Latal and W. Schweiger (Springer-Verlag, Berlin, 1994). }}}
 \end{center}
\vskip .1in
\begin{center}

Keith A. Olive

{\it School of Physics and Astronomy, University of Minnesota}

{\it Minneapolis, MN 55455, USA}

\vskip .2in

\end{center}

\baselineskip=15pt

\def\beq{\begin{equation}}
\def\eeq{\end{equation}}
\def\bl{B - L}
\def\la{~\mbox{\raisebox{-.6ex}{$\stackrel{<}{\sim}$}}~}
\def\ga{~\mbox{\raisebox{-.6ex}{$\stackrel{>}{\sim}$}}~}
\section{The Standard Model}

One of the most basic questions we can ask about the Universe is:
What is the origin of matter? There are of course many ways in
which to interpret this question, and there are varying depths
to which it can  be answered. Essentially all of the mass in the Universe
that is {\em observed} is in the form of baryons, today consisting
of protons and neutrons in the nuclei of atoms. While baryons may not be
the dominant component of the Universe, they are without a doubt present
and essential to our existence. However, the fact that a significant part
of the mass of the Universe is baryonic is not in  and of itself
surprising.  The lightest baryons are relatively
long lived by particle physics
standards and massive. Protons have extremely long lifetimes
(or, in a boring world may be stable) and neutrons live long
enough to become incorporated primarily into helium nuclei in the early
Universe (see below). While electrons are stable (so long as electric
charge is conserved), and they are present in numbers equal to that
of protons, they are too light to make a significant contribution to the
mass density of the Universe. Other stable particles which may yet
be found  to be massive, such as neutrinos, or still to be discovered
such as the lightest supersymmetric particle, may in fact dominate the
overall mass density of the Universe. There are however,
two known particles
which on the basis of mass and lifetime could be expected to
contribute to the mass of the Universe: the anti-proton and the anti-neutron.
${\bar p}$ and ${\bar n}$ have, of course, exactly the same mass and lifetime
as $p$ and $n$.  Yet these antibaryons are not observed in any abundance
in nature.  The creation of this asymmetry between baryons and
anti-baryons or between matter and  anti-matter is the subject of these
lectures.

To deal with the specific problem of the baryon asymmetry, it will be
useful to briefly review the standard cosmological model as a
framework towards a solution. To put the problem in perspective, it useful
to have an idea of the general sequence of events which are
believed  to have occurred since the big bang.  The earliest times (after
the big bang) that we are able to discuss are after $t \simeq 10^{-44}$s,
or at temperatures of about $10^{18}$ GeV. This period is the Planck
epoch and a description of events at or prior to this time would
require a more complete theory of quantum gravity which may yet
be found in string theory.  The Grand Unified (GUT) scale is typically
at $T \sim 10^{15}$ GeV at times of about $10^{-35}$s. Standard models
of baryogenesis and inflation may have played important roles at this time.
Barring new interactions at an intermediate scale, electroweak symmetry
breaking then occurred at times of order $10^{-10}$s at the electroweak
scale of 100 GeV.  Quark-gluon confinement should have taken place
at $t \sim 10^{-5}$s at $T \sim \Lambda_{\rm QCD} \sim 100$ MeV.
Big bang nucleosynthesis and the formation of the light element
isotopes of D, $^3$He, $^4$He and $^7$Li took place between
1 and 100 s, at temperatures below 1 MeV.  It wasn't until
$t \sim 10^{12}$ s or $T \sim 1$ eV that recombination of neutral
hydrogen occurred and the formation of galaxies began.
Finally to put things in perspective, the age of the Universe today is
$ \sim 10^{17}$ s and the temperature is the well known 2.726 K
as measured by COBE \cite{cobet}

The standard big bang model assumes homogeneity and
isotropy, so that space-time can be described by the
Friedmann-Robertson-Walker metric which in co-moving coordinates is given by
\beq
	ds^2  = -dt^2  + R^2(t)\left[ {dr^2 \over \left(1-kr^2\right) }
      + r^2 \left(d\theta^2  + \sin^2 \theta d\phi^2 \right)\right]
\label{met}
\eeq
where $R(t)$ is the cosmological scale factor and $k$ is the three-space
curvature constant ($k = 0, +1, -1$ for a spatially flat, closed or open
Universe). $k$ and $R$ are the only two quantities in the
metric which distinguish it from flat Minkowski space.
It is  also common to assume
 the perfect fluid form for the energy-momentum
tensor
\beq
	T_{\mu\nu}   = pg^{\mu\nu}   + (p + \rho)u^\mu u^\nu
\eeq
where $g_{\mu\nu}$   is the space-time metric described by (\ref{met}),
 $p$ is the isotropic
pressure, $\rho$ is the energy density and $u^\mu  = (1,0,0,0)$
 is the velocity vector
for the isotropic fluid.  Einstein's equation  yield the
Friedmann equation,
\beq
	H^2  \equiv \left({\dot{R} \over R}\right)^2  = {1 \over 3} 8 \pi G_N \rho
 - { k \over R^2}  + {1 \over 3} \Lambda
\label{H}
\eeq
and
\beq
	\left({\ddot{R} \over R}\right) = {1 \over 3} \Lambda -
 {1 \over 6} 8 \pi G_N ( \rho + 3p)
\eeq
where $\Lambda$ is the cosmological constant,
or equivalently from $	{T^{\mu\nu}}_{;\nu}   =  0$
\beq
	\dot{\rho} = -3H(\rho + p)
\label{rhod}
\eeq
These equations form the basis of the standard big bang model.

At early times ($t < 10^5 $ yrs) the Universe is thought to have been
dominated by radiation so that the equation of state can be given by $p =
\rho/3$.  If we neglect the contributions to $H$ from $k$ and $\Lambda$
 (this is always a
good approximation for small enough $R$) then we find that
\beq
	R(t) \sim t^{1/2} \qquad
\eeq
and $\rho \sim R^{-4}$  so that $t \sim (3/32 \pi G_N\rho)^{1/2}$.
  Similarly for a matter or dust
dominated Universe with $p = 0$,
\beq
	R(t) \sim t^{2/3}
\eeq
and $\rho \sim R^{-3}$.  The Universe makes the transition
 between radiation and matter
domination when $\rho_{rad} = \rho_{matter}$ or
when $T \simeq$ few $\times~10^3$ K.

In the absence of a cosmological constant, one can
define a critical energy density $\rho_c$
  such that $\rho =\rho_c$  for $k = 0$
\beq
	\rho_c  = 3H^2 / 8 \pi G_N
\eeq
In terms of the present value of the Hubble parameter this is,
\beq
	\rho_c  = 1.88 \times 10^{-29} {h_o}^2  {\rm g cm}^{-3}
\eeq
where
\beq
	h_o  = H_o /(100 {\rm km Mpc}^{-1}   {\rm s}^{-1}  )
\eeq
The cosmological density parameter is then defined by
\beq
	\Omega \equiv {\rho \over \rho_c}
\eeq
in terms of which the Friedmann equation, Eq. (\ref{H}), can be rewritten as
\beq
	(\Omega - 1)H^2  = {k \over R^2}
\label{o-1}
\eeq
so that $k = 0, +1, -1$ corresponds to $\Omega = 1, \Omega > 1$
 and $\Omega < 1$.
Observational limits on $h_o$ and $\Omega$ are\cite{tonry}
\beq
0.4 \le h_o \le 1.0 \qquad 0.1 \le \Omega \le 2
\label{range}
\eeq

It is important to note that $\Omega$ is a function of
time or of the scale factor.
The evolution of $\Omega$ is shown in figure 1 for $\Lambda = 0$.

\begin{figure}
\vspace{7.5cm}
\caption{The evolution of the cosmological density parameter, $\Omega$,
as a function of the scale factor for a closed, open and
spatially flat Universe.}
\end{figure}

\noindent For a spatially flat Universe, $\Omega = 1$ always. When $k = +1$,
there is a maximum value for the scale factor $R$. At early
times (small values of $R$), $\Omega$ always tends to one.
Note that the fact that we do not yet know the sign of
$k$, or equivalently whether $\Omega$ is larger than or smaller than
unity, implies that we are at present still at the very left in the
figure.  What makes this peculiar is that one would normally expect that
the sign of $k$ to become apparent after a Planck time of $10^{-43}$ s.
It is extremely puzzling that some $10^{60}$ Planck times later,
we still do not know the sign of $k$.

\subsection{The Hot Thermal Universe}

 The epoch of recombination occurs when electrons
 and protons form neutral hydrogen  through
$e^{ -  } +  p     \rightarrow $  H  $+   \gamma $
  at a temperature
$T_{ R} { \sim }$  few $\times 10^{3}$ K  ${ \sim}1$ eV.  For $T < T_{R}$,
photons are decoupled while for $T > T_{ R}$,  photons are
 in thermal equilibrium and the Universe is usually
 taken to be radiation dominated so that the content
 of the radiation plays a very important role.  Today, the content
 of the microwave background consists of photons with
$T_o =  2.726  \pm .01$ K\cite{cobet}.
We can calculate the energy density of photons from
\beq
 \rho_\gamma  = \int E_\gamma dn_\gamma
\label{rhog}
\eeq
 where the density of states is given by
\beq
dn_\gamma  =   {g_\gamma \over 2
 \pi^{ 2}}[exp(E_\gamma/T)-1]^{ -  1} q^{ 2} dq
\eeq
 and $g_\gamma = $  2 simply counts the number of degrees of freedom
 for photons,
$E_\gamma =  q$ is just the photon energy (momentum).
 (I am using units such that
$\hbar =  c  = k_{ B}   =$  1 and will do so through the remainder
 of these lectures.)
Integrating (\ref{rhog}) gives
\beq
\rho_\gamma = {\pi^2 \over 15} T^4
\eeq
 which is the familiar blackbody result.

In general, at very early times, at very high temperatures,
 other particle degrees of freedom join the radiation background when
$T{ \sim } m_{i}$  for each  particle type i if that type is brought
 into thermal equilibrium through interactions.  In equilibrium
the energy density of a particle type i is given by
\beq
 \rho_{i}  = \int E_{i} dn_{q_{i}}
\eeq
 and
\beq
 dn_{q_{i}} = {g_{i} \over 2  \pi^{ 2}}[exp[(E_{q_{i}} -
\mu_{i})/T] \pm 1]^{ -1 }q^{2}dq
\eeq
where again $g_{i}$ counts the total number of degrees of freedom for type i,
\beq
 E_{q_{i}} =  \left(m_{i}^{2} + q_{i}^{ 2}\right)^{1/2}
\eeq
$\mu_{i}$ is the chemical potential if present and  $ \pm$
corresponds to either Fermi or Bose statistics.

In the limit that  $T \gg m_{i} $  the total energy density can
 be conveniently expressed by
\beq
 \rho {} = \left( \sum_B g_{B} + {7 \over 8} \sum_F  g_{F} \right)
   {\pi^{ 2} \over 30}  T^{4}     \equiv    {\pi^{ 2} \over 30} N(T) T^{4}
\label{NT}
\eeq
 where $g_{B(F)} $  are the total number of boson (fermion)
degrees of freedom and the sum runs over all boson (fermion) states with
$m \ll T$.  The factor of 7/8 is due to the difference between
 the Fermi and Bose integrals.  Equation (\ref{NT}) defines N(T)
 by taking into account  new particle degrees
of freedom as the temperature is raised.

In the radiation dominated epoch,
eq. (\ref{rhod}) can be integrated (neglecting the $T$-dependence of $N$)
giving us a relationship between
 the age of the Universe and its temperature
\beq
 t = \left({90 \over 32 \pi^3 G_{ N} N(T)}\right)^{ 1/2}  T^{ -  2}
\eeq
 Put into a more convenient form
 \beq
 tT_{ MeV}^{ 2}  =
2.4 [N(T)]^{ -  1/2}
\eeq
 where t is measured in seconds and
$T_{ MeV} $  in units of MeV.

 The value of $N(T)$ at any given temperature depends
 on the particle physics model.  In the standard $SU(3) \times
SU(2)\times U(1)$  model, we can specify $N(T)$ up to
temperatures of 0(100) GeV.
  The change in N can be seen in the following table.

\begin{table}
\caption{Effective numbers of degrees of freedom in the standard model.}
\vspace{2pt}
\begin{tabular}{llc}
\hline
Temperature & New Particles \qquad
&$4N(T)$ \\
\hline\rule{0pt}{12pt}
$T < m_{ e}$   &     $\gamma$'s +   $\nu$'s & 29 \\
$m_{ e} <   T  < m_\mu$ &    $e^{\pm}$ & 43 \\
$m_\mu <  T  < m_\pi$  &   $\mu {}^{\pm}$ & 57 \\
$m_\pi <  T < T{ c}^{*}$  & $\pi$'s & 69 \\
$T_{ c} <  T  < m_{\rm strange~~~~~~~}$ \qquad &
  -  $\pi$'s + $  u,{\bar u},d,{\bar d}$ + gluons &  205 \\
$m_{ s} <  T < m_{ charm}$ & $s,{\bar s}$ & 247 \\
$m_{ c} <  T < m_\tau$ &  $c,{\bar c}$ & 289 \\
$m_\tau < T < m_{ bottom}$ & $\tau {}^{\pm}$ & 303 \\
$m_{ b} < T < m_{ W,Z}$ & $b,{\bar b}$ & 345 \\
$m_{ W,Z} <  T < m_{ top}$ & $W^{\pm}, Z$ & 381 \\
$ m_t < T < m_{Higgs}$ & $t,{\bar t}$ & 423 \\
$M_H < T $ & $H^o$ & 427 \\
\hline
\end{tabular}
\newline
*$T_{ c}$
corresponds to the confinement-deconfinement transition between
 quarks and hadrons. $g(T)
= N(T)$  is shown in Figure 2 for $T_{ c}  =  150$ and $400$ MeV.
It has been assumed that $m_{Higgs} > m_{top}$.
\end{table}

 At higher temperatures, $N(T)$ will be model dependent.
  For example, in the minimal $SU(5)$ model, one needs to add
 to $N(T)$, 24 states for the X and Y gauge bosons,
another 24 from the adjoint Higgs, and another 6 (in addition
to the 4 already counted in $W^\pm, Z$ and $H$) from the 5
of Higgs.
  Hence for
$T > M_{ X}  $  in minimal $SU(5)$, $N(T)  =  160.75$.
 In a supersymmetric model this would at least double,
with some changes possibly necessary in the table if the
 lightest supersymmetric particle has a mass below
$M_{ H}.$

\begin{figure}
\vspace{7.5cm}
\caption{The effective numbers of relativistic degrees of freedom
as a function of temperature. The dashed lines correspond to free
quarks and hadrons.}
\end{figure}

The notion of equilibrium also plays an important role in the
standard big bang model. If, for example, the Universe were not
 expanding, then given enough time, every particle state would
come into equilibrium with each other.  Because of the expansion
of the Universe, certain rates might be too slow indicating,
for example, in a scattering process that the two incoming states
 might never find each other to bring about an interaction.
Depending on their rates, certain interactions may pass in and
 out of thermal equilibrium  during the course of the Universal expansion.
 Quantitatively, for each particle i, we will require that some rate
$  \Gamma {}_{ i} $  involving that type be larger than the expansion
 rate of the Universe or
\beq
  \Gamma {}_{ i}  > H
\eeq
  in order to be in thermal equilibrium.

A good example for a  process in equilibrium at some stage
 and out of equilibrium at others is that of neutrinos.
If we consider the standard neutral or charged-current interactions such as
$e^{ +} + e^{ -  }     \leftrightarrow {}   \nu {}  +  \bar \nu $
or $e +   \nu {}    \leftrightarrow {}  e  +   \nu {}$  etc.,
 very roughly the rates for these processes will be
 \beq
 \Gamma {} =  n \langle \sigma v \rangle
 \eeq
 where  $  \langle \sigma v \rangle $
is the thermally averaged  weak interaction cross section
\beq
 \langle \sigma v \rangle { \sim }~0(10^{ -  2}) T^{ 2} /M_{ W}^4
\eeq
and n is the number density of leptons.
 Hence the rate for these interactions is
\beq
 \Gamma {}_{\rm wk}   { \sim  }~0(10^{ -  2}) T^{ 5}/M_{ W}^4
\eeq
The expansion rate, on the other hand, is just
\beq
  H  =  \left({8 \pi {}G_{ N}  \rho {} \over 3}\right)^{ 1/2}
 =  \left({8  \pi {}^{ 3} \over 90}  N(T) \right)^{ 1/2}  T^{ 2}/M_{ P}
 \sim 1.66 N(T)^{ 1/2}  T^{ 2}/M_{ P}.
\eeq
The Planck mass $M_{ P} = G_N^{-1/2} =  1.22 \times 10^{19}$ GeV.

Neutrinos will be in equilibrium when  $  \Gamma {}_{\rm wk} >  H $ or
\beq
T > (500 M_{ W}^4)/M_{ P})^{ 1/3}  { \sim }~1 MeV .
\eeq
The temperature at which these rates are equal
 is commonly referred to as the decoupling or freeze-out
temperature and is defined by
\beq
  \Gamma(T_{ d}) = H(T_{ d})
\eeq
 For temperatures $T > T_{ d}$,
neutrinos will be in equilibrium, while for $T < T_{ d }$
 they will not. Basically, in terms of their interactions,
the expansion rate is just too fast and they never
{\em ``see"}  the rest of the matter in the Universe (nor themselves).
  Their momenta will simply redshift and their effective temperature
(the shape of their momenta distribution is not changed from that
of a blackbody) will simply fall with
$T { \sim  } 1/R.$

\subsection{Big Bang Nucleosynthesis}
An essential element of the standard cosmological model is
big bang nucleosynthesis, the theory which predicts the abundances
of the light element isotopes D,$^3$He, $^4$He, and $^7$Li.
As was mentioned earlier, nucleosynthesis takes place at a
temperature scale of order 1 MeV. At temperatures above 1 MeV,
the weak interactions, being in equilibrium, determined the
ratio of neutrons to protons.  Near 1 MeV, these interactions:
$n + e^+ \leftrightarrow p + {\bar \nu}$; $n + \nu \leftrightarrow
p + e^-$; and $n \leftrightarrow p + e^- + {\bar \nu}$; as the $e,\nu$
interactions discussed above, drop out of equilibrium. Although
the binding energy of deuterium is 2.2 MeV, due to the high photon
to baryon ratio ($10^{10}$), nucleosynthesis is delayed until
about $T \sim 2.2/\ln 10^{10} \sim 0.1$ MeV, when
deuterium can be formed without significant dissociation.
Afterwhich, nucleosynthesis proceeds rapidly with the build-up of the
light elements.

\begin{figure}
\vspace{20cm}
\caption{The abundances of the light elements as a function
 of the baryon-to-photon ratio.}
\end{figure}

The nuclear processes lead primarily to $^4$He, which is produced at about
24\% by mass. Lesser amounts of the other light elements are produced:
about $10^{-5}$ of D and $^3$He and about $10^{-10}$ of $^7$Li
by number relative to H. The abundances of the light
elements depend almost solely
on one key parameter, the baryon-to-photon ratio, $\eta$.  In figure 3,
(taken from ref.\cite{wssok} )
the predicted abundances of the light elements are shown as  a function
of $\eta_{10} = 10^{10} \eta$. In figure 3, the boxes
correspond to acceptable
values for the abundances as determined from the
observations. The band for the
$^4$He curve shows the sensitivity to the
neutron half life.  The vertical band
shows the overall range of $\eta$ in which agreement is achieved between
theory and observation for all of the light elements.
{}From the figure we see that consistency is found for
\beq
2.8 \times 10^{-10} < \eta < 3.2 (4) \times 10^{-10}
\label{etas}
\eeq
where the bound can be as high as $\eta_{10} < 4$ when the uncertainties in
$^7$Li cross-sections are accounted for.

It is important to note that $\eta$ is related to the fraction of $\Omega$
contained in baryons, $\Omega_B$
\beq
\Omega_B = 3.66 \times 10^7 \eta h_o^{-2} (T_o/2.726 )^3
\eeq
Using the limits on $\eta$ and $h_o$, one finds that $\Omega_B$
is restricted to  a range
\beq
0.01 < \Omega_B < 0.08
\eeq
and one can conclude that the Universe is not closed by baryons.
This value of $\eta$ is the one that we try to explain by
big bang baryogenesis.

\subsection{Problems with the (Non-Inflationary) Standard Model}

Despite the successes of the standard big bang model, there are a
number of unanswered questions that appear difficult to explain without
imposing what may be called unnatural initial conditions.
The resolution of these problems may lie in a unified theory of gauge
interactions or possibly in a theory which includes gravity.
For example, prior to the advent of grand unified theories (guts),
the baryon-to-photon ratio, could have been viewed as
being embarrassingly small.
Although, we still do not know the
 precise mechanism for generating the baryon asymmetry, many
quite acceptable models
are available as will be discussed in some detail
 for the better part of these lectures.
In a similar fashion, it is hoped that a field theoretic
description of inflation may resolve the problems outlined below.

	\subsubsection{The Curvature Problem}
	The bound on $\Omega$ in Eq. (\ref{range}) is
curious in the fact that at the present
time we do not know even the sign of the curvature term in the Friedmann
equation (\ref{H}), i.e., we do not know if the Universe is open, closed or
spatially flat.

	The curvature problem (or flatness problem or age problem)
can manifest itself in
several ways. For a radiation
dominated gas, the entropy density $s \sim T^3$  and $R \sim T^{-1}$.
 Thus assuming an adiabatically expanding
Universe, the quantity ${\hat k} = k/R^2T^2$ is a dimensionless constant.
  If we now
apply the limit in Eq. (\ref{range}) to Eq. (\ref{o-1}) we find
\beq
{\hat	k} = {k \over R^2 T^2}  = {(\Omega_o  - 1){H_o}^2 \over {T_o}^2}
  < 2 \times 10^{-58}
\eeq
This limit on $k$ represents an initial condition on the cosmological model.
The problem then becomes what physical processes in the early Universe
produced a value of $\hat{k}$ so extraordinarily close to zero
(or $\Omega$ close to one).
	A more natural initial condition might have
been $\hat{k} \sim 0(1)$.  In this
case the Universe would have become curvature
dominated at $T \sim 10^{-1} M_P$.
  For
$k = +1$, this would signify the onset of recollapse.
As already noted earlier, one would naturally
expect the effects of curvature
(seen in figure 1 by the separation of the
three curves) to manifest themselves at times on the order
of the Planck time as gravity  should provide the only
dimensionful scale in this era.
If we view the evolution of $\Omega$ in figure 1
as a function of time, then the it would appear that the time
$t_o = 13$ Gyr = $7.6 \times 10^{60} {M_P}^{-1}$ ($\sim$ the current
 age of the Universe) appears at the far left of x-axis, ie before the
curves separate. Why then has the Universe lasted so long before revealing
the true sign of $k$?

  Even for $\hat{k}$ as small as
$0(10^{-40})$ the Universe would have become curvature
 dominated when $T \sim 10$ MeV,
ie, before the onset of big bang nucleosynthesis.  Of course,
it is also possible that $k = 0$ and the Universe is actually spatially flat.
In fact, today we would really expect only one of two possible
values for $\Omega$: 0 or 1.

\subsubsection{The Horizon Problem}

	Because of the cosmological principle, all physical length scales grow
as the scale factor $R(t) \sim t^{2/3\gamma}$, with $\gamma$ defined by
$p = (\gamma - 1) \rho$.
 However, causality implies the existence
of a particle (or causal) horizon $d_H(t) \sim t$,which
  is the maximal physical distance light can travel from the co-
moving position of an observer at some initial time ($t=0$) to time $t$.
 For $\gamma > {2 \over 3}$,
 scales originating outside of the horizon will eventually
become part of our observable Universe.  Hence we would expect to see
anisotropies on large scales \cite{hor}.

	In particular, let us consider the microwave background today.  The
photons we observe have been decoupled
since recombination at $T_d \sim 4000$K.
At that time, the horizon volume was simply $V_d \propto {t_d}^3$,
 where $t_d$  is the age of
the Universe at $T = T_d$.
Then $t_d  = t_o (T_o/T_d)^{3/2}    \sim 2 \times 10^5$ yrs, where
$T_o =  2.726$K\cite{cobet} is the present temperature of the
microwave background.
 Our present horizon
volume $V_o \propto {t_o}^3$  can be scaled back to
 $t_d$ (corresponding to that part of the
Universe which expanded to our present visible
Universe) $V_o(t_d) \propto V_o
(T_o/T_d)^3$.  We can now compare $V_o(t_d)$ and $V_d$.  The ratio
\beq
	{V_o(t_d) \over V_d}
 \propto {V_o {T_o}^3 \over V_d {T_d}^3}
  \propto {{t_o}^3 {T_o}^3 \over {t_d}^3 {T_d}^3}  \sim 5 \times 10^4
\eeq
corresponds to the number of horizon volumes or casually distinct regions at
decoupling which are encompassed in our present visible horizon.

	In this context, it is astonishing that the microwave
 background appears highly
isotropic on large scales with
 $\Delta T/T = 1.1 \pm 0.2 \times 10^{-5}$
at angular separations of $10^o$\cite{cobe}.  The horizon
problem, therefore, is the lack of an explanation as to why nearly $10^5$
causally disconnected regions at recombination all had the same temperature
to within one part in $10^{-5}$.

\subsubsection{Density Perturbations}

	Although it appears that the Universe is extremely isotropic and
homogeneous on very large scales
(in fact the standard model assumes complete isotropy and
homogeneity) it is very inhomogeneous on small scales.  In other words,
there are planets, stars, galaxies, clusters, etc.
 On these small scales there
are large density perturbations namely $\delta \rho/\rho \gg 1$.
At the same time, we know from the isotropy of the microwave background
that on the largest scales, $\delta \rho /\rho \sim 3 \Delta T/ T
\sim O(10^{-5})$ \cite{cobe} and these perturbations must have grown to
$\delta \rho/\rho \sim 1$ on smaller scales.

	In an expanding Universe, density perturbations evolve with time\cite{peeb}.
 The evolution of  the Fourier transformed quantity
 ${\delta \rho \over \rho}(k,t)$ depends on
the relative size of the wavelength $\lambda \sim k^{-1}$
 and the horizon scale $H^{-1}$.  For
$k \ll H$, (always true at sufficiently early times)
$\delta \rho/\rho \propto t$ while for $k \gg H$,
$\delta \rho/\rho$ is $\simeq$ constant (or grows moderately
as $\ln t$)
 assuming a radiation dominated Universe.
In a matter dominated Universe, on scales larger than the Jean's
length scale (determined by $k_J = 4\pi G_N \rho_{matter}/{v_s}^2$, $v_s$ =
sound speed) perturbations grow with the scale factor $R$.
  Because of the
growth in $\delta \rho/\rho$, the microwave background
limits force $\delta \rho/\rho$ to be extremely
small at early times.

	Consider a perturbation with wavelength on the order of a galactic
scale.  Between the Planck time and recombination, such a perturbation would
have grown by a factor of $O(10^{57})$ and the anisotropy limit
of $\delta \rho/\rho \la 10^{-5}$
implies that $\delta \rho/\rho < 10^{-61}$ on the scale
 of a galaxy at the Planck time.  One
should compare this value with that predicted from purely random (or
Poisson) fluctuations of $\delta \rho/\rho \sim 10^{-40}$
   (assuming $10^{80}$  particles (photons) in
a galaxy) \cite{bg}.  The extent of this limit
is of course related to the fact
that the present age of the Universe is so great.

	An additional problem is related to the formation time of the
perturbations.  A perturbation with a wavelength large enough to correspond
to a galaxy today must have formed with wavelength modes much greater than
the horizon size if the perturbations are primordial as is generally
assumed.  This is due to the fact that the wavelengths red shift as $\lambda
\sim R \sim t^{1/2}$ while the horizon size grows linearly.
It appears that a mechanism for generating
perturbations with acausal wavelengths is required.

\subsubsection{The Magnetic Monopole Problem}

	In addition to the much desired baryon asymmetry produced by grand
unified theories, a less favorable aspect is also present; guts predict the
existence of magnetic monopoles.  Monopoles will be produced \cite{thp}
whenever any simple group [such as $SU(5)$] is broken down to a gauge group
which contains a $U(1)$ factor [such as $SU(3) \times SU(2) \times U(1)$].
 The mass of
such a monopole would be
\beq
	M_m \sim M_{gut}/\alpha_{gut}  \sim 10^{16}  GeV.
\eeq
The basic reason monopoles are produced is that in the breaking of SU(5) the
Higgs adjoint needed to break SU(5)
cannot align itself over all space \cite{kib}.
 On scales larger than the
horizon, for example, there is no reason to expect the direction of the
Higgs field to be aligned.  Because of this randomness, topological knots
are expected to occur and these are the magnetic monopoles.  We can then
estimate that the minimum number of monopoles produced \cite{mon}
 would be roughly
one per horizon volume or causally connected region at the time of the $SU(5)$
phase transition $t_c$
\beq
	n_m  \sim (2t_c)^{-3}
\eeq
resulting in a monopole-to-photon ratio expressed in terms of the transition
temperature of
\beq
	{n_m \over n_\gamma} \sim \left({10T_c \over M_P}\right)^3
\label{mop}
\eeq
	The overall mass density of the Universe can be used to place a
constraint on the density of monopoles.
For $M_m  \sim 10^{16}$ GeV and $\Omega_m {h_o}^2  \la 1$ we have that
\beq
	{n_m \over n_\gamma}  \la 0(10^{-25})
\eeq
The predicted density, however, from (\ref{mop}) for $T_c \sim M_{gut}$
\beq
	{n_m \over n_\gamma}  \sim 10^{-9}
\eeq
Hence, we see that standard guts and cosmology have a monopole problem.

\subsection{Inflation}
All of the problems discussed above can be neatly resolved if the
Universe underwent a period of cosmological inflation \cite{infl}.
 During a phase transition, our assumptions of an
adiabatically expanding universe may not be valid.
If we look at a scalar potential
describing a phase transition from a symmetric false vacuum state
 $\langle \phi \rangle  = 0$ for some
scalar field $\phi$ to the broken true vacuum at
 $\langle \phi \rangle  = v$ as in figure 4,
and suppose we find that upon solving the equations of motion for the
scalar field that the field evolves slowly from the symmetric
state to the global minimum (this will depend on the details of the
potential).  If the evolution is slow enough, the universe may become dominated
by the vacuum energy density associated with
the potential near $\eta \approx 0$.
The energy density of the symmetric vacuum, V(0) acts as a cosmological
constant with
\beq
	\Lambda = 8\pi V(0) {M_P}^2
\eeq
During this period of slow evolution, the energy
density due, to say, radiation will fall below the vacuum
energy density, $\rho \ll V(0)$.  When this happens, the expansion
rate will be dominated by the constant V(0) and from Eq. (\ref{H})
we find an exponentially expanding solution
\beq
R(t) \sim e^{\sqrt{\Lambda/3}~t}
\label{DS}
\eeq
When the field evolves towards the global minimum
it will begin to oscillate
about the minimum, energy will be released
during its decay and a hot thermal universe will be restored.
 If released
fast enough, it will produce radiation at a temperature
${T_R}^4 \la V(0)$.  In this reheating process entropy has been created and
	$(RT)_f  > (RT)_i $.
Thus we see that during a phase
 transition the relation $RT \sim$ constant, need not hold true and thus
our dimensionless constant $\hat{k}$ may actually not have been constant.

\begin{figure}
\vspace{13cm}
\caption{A typical potential suitable for the inflationary universe
scenario.}
\end{figure}

If during the phase transition,  the value of
 $RT$ changed by a factor of $0(10^{29})$, the cosmological
 problems would be solved.  The isotropy would in a sense be
 generated by the immense expansion; one small causal region
 could get blown up and hence our entire visible Universe would
have been at one time in thermal contact.  In addition,
the parameter $\hat{k}$ could have started out $0(1)$ and have
 been driven small by the expansion. Density perturbations
will be stretched by the expansion, $\lambda \sim R$.
Thus it will appear that $\lambda \gg H^{-1}$ or that the perturbations
have left the horizon.  It is actually just that the size of the causally
connected region is now no longer simply $H^{-1}$. However, not only does
inflation offer an explanation for large scale perturbations, it also
offers a source for the perturbations themselves \cite{pert}.
Monopoles would also be diluted away.

The cosmological problems could be solved if
\beq
	H\tau > 65
\eeq
where $\tau$ is the duration of the phase
transition, density perturbations are
produced and do not exceed the limits imposed by
the microwave background anisotropy,  the vacuum energy density was converted
 to radiation so that the reheated temperature is sufficiently
high, and baryogenesis is realized.

For the purposes of discussing baryogenesis, it will be sufficient to
consider only a generic model of inflation whose potential
is of the form:
\begin{equation}
 V( \eta ) = {{\mu}^4} P( \eta )
\label{a}
\end{equation}
where $\eta$ is the scalar field driving inflation, the inflaton,
$\mu$ is an as yet unspecified mass parameter, and $P(\eta)$ is a
function of $\eta$ which possesses the features necessary for
inflation, but contains no small parameters.
I.e. $P(\eta)$ takes the form,
\beq
P(\eta) = P(o) + m^2 \eta^2 + \lambda_3 \eta^3 + \lambda_4 \eta^4   +...
\eeq
where all of the couplings in $P$ are $O(1)$ and $...$ refers to
possible non-renormalizable terms. Most of the useful
inflationary potentials can be put into the form of Eq. (\ref{a}).

The requirements for successful inflation boil down to: 1) enough inflation;
\beq {\partial^2 V \over \partial \eta^2}\mid_{\eta \sim {\eta_i} \pm H}
< {3 H^2 \over 65} = {8 \pi V(0) \over 65 {M_P}^2}
\eeq
2) density perturbations of the right magnitude\cite{pert};
\beq
{\delta \rho \over \rho} \simeq {H^2 \over 10 \pi^{3/2} \dot{\eta}}
\simeq \left\{ \begin{array}{l}
\left({32 \lambda_4 \over 3 \pi^3}\right)^{1/2}
{1 \over 10} \ln^{3/2}(H k^{-1}) {\mu^2 \over {M_P}^2} \\
\left({\lambda_3 \over H \pi^{3/2}}\right)
{1 \over 10} \ln^{2}(H k^{-1}) {\mu^4 \over {M_P}^4} \\
\left({8  \over 3 \pi^2}\right)^{1/2}
{1 \over 10} {m \over M_P} \ln (H k^{-1})
{\mu^2 \over {M_P}^2} \end{array} \right.
\label{perts}
\eeq
given here for scales which ``re-enter" the
 horizon during the matter dominated
era. These reduce approximately to
\beq
{\delta \rho \over \rho} \sim O(100) {\mu^2 \over {M_P}^2}
\label{drho}
\eeq
3) baryogenesis; the subject of the remaining lectures.

       For large scale fluctuations of the type measured by COBE\cite{cobe},
we can use Eq. (\ref{drho}) to fix the inflationary scale $\mu$.
 The magnitude of the
density fluctuations can be related to the observed quadrupole\cite{pee}
moment:
\begin{equation}
{\langle{a_2^2}\rangle = \frac{5}{6}
2{\pi^2}{(\frac{\delta\rho}{\rho})^2}}
\end{equation}
 The
observed quadrupole moment gives \cite{cobe}:
\begin{equation}
{\langle{a_2^2}\rangle = (4.7\pm2)\times{10^{-10}}}
\end{equation}
or
\begin{equation}
{\frac{\delta\rho}{\rho} = (5.4\pm1.6)\times{10^{-6}}}
\end{equation}
which in turn fixes the coefficient $\mu$ of the inflaton potential\cite{cdo}:
\begin{equation}
{\frac{\mu^2}{M_P^2} = {\rm few} \times{10^{-8}}}
\label{cobemu}
\end{equation}

       Fixing $({\mu^2}/{M_P^2})$ has immediate general consequences
for inflation\cite{eeno}. For example, the Hubble parameter during inflation,
${{H^2} \simeq (8\pi/3)({\mu^4}/{M_P^2})}$ so that $H \sim
10^{-7}M_P$. The duration of inflation is $\tau \simeq
{M_P^3}/{\mu^4}$, and the number of e-foldings of expansion is $H\tau
\sim 8\pi({M_P^2}/{\mu^2}) \sim 10^{9}$. If the inflaton decay rate
goes as $\Gamma \sim {m_{\eta}^3}/{M_P^2} \sim {\mu^6}/{M_P^5}$, the
universe recovers at a temperature $T_R \sim (\Gamma{M_P})^{1/2} \sim
{\mu^3}/{M_P^2} \sim 10^{-11} {M_P} \sim 10^8 GeV$. Recall that
before COBE all that could be set was an upper limit on $\mu$.

\section{Big Bang Baryogenesis}
It appears
that there is apparently very little antimatter
in the Universe and that the number of photons greatly exceeds
 the number of baryons.
 In the standard model, the entropy density today is related to
$n_\gamma {}$  by
\beq
s  \sim 7n_\gamma
\eeq
so that eq. (\ref{etas}) implies $n_{ B}/s  \sim 4 \times 10^{-11}$.
In the absence of baryon number violation or entropy production
this ratio is conserved however and hence represents a potentially undesirable
initial condition.

Let us for the moment, assume that in fact  $\eta  = $  0.  We can
compute the final number density of nucleons
left over after annihilations of baryons and
antibaryons  have frozen out.  At very high
temperatures (neglecting a quark-hadron
 transition) $T
>$  1 GeV, nucleons were in thermal equilibrium with
the photon background and $n
_{B} = n_{\bar B} = (3/2)n_\gamma$  (a factor of 2 accounts
 for neutrons and protons and the factor 3/4
 for the difference between fermi and bose statistics).
 As the temperature fell below $m_N$
  annihilations kept the nucleon density at its equilibrium value
$(n_B/n_\gamma) = (m_{N}/T)^{3/2} {\rm exp}(-m_{N}/T)$
 until the annihilation rate
$\Gamma_A \simeq n_B m_\pi^{-2}$
 fell below the expansion rate. This occurred at $T
\simeq$  20 MeV.  However, at this time the nucleon
number density had already dropped to
\beq
n_B/n_\gamma = n_{\bar B}/n_\gamma \simeq 10^{-18}
\eeq
 which is eight orders of magnitude too small \cite{Gary} aside from
the problem of having to separate the baryons from the antibaryons.
 If any separation did occur at higher temperatures
(so that annihilations were
as yet incomplete) the maximum distance scale on which separation could occur
is the causal scale related to the age of the Universe at that time.  At $T
=$  20 MeV, the age of the Universe was only $t  =  2 \times 10^{-3}$
sec.  At that time, a causal region (with distance scale defined by 2ct)
could only have contained
$10^{-5} M_\odot$  which is very far from the galactic mass scales
which we are asking for separations to occur,
$10^{12} M_\odot$. In spite of all of these problems, $\eta = 0$, implies that
the Universe as a whole is baryon symmetric, thus unless baryons are
separated on extremely large (inflationary) domains, in which case we might
just as well worry again about $\eta \ne 0$, there should be
antimatter elsewhere in the Universe.  To date, the only antimatter
observed is the result of a high energy collision, either in an
accelerator or in a cosmic-ray collision in the atmosphere. There has been no
sign to date of any primary antimatter, such as an
anti-helium nucleus ${\bar \alpha}$
found in cosmic-rays.

\subsection{The out-of-equilibrium decay scenario}

The production of a net baryon asymmetry requires baryon number violating
interactions, C and CP violation and a departure
from thermal equilibrium\cite{sak}.
The first two of these ingredients are contained in guts,
the third can be realized in an expanding universe
 where as we have seen, it is not uncommon that interactions
come in and out of equilibrium.
In SU(5), the fact that quarks and leptons are in the same multiplets allows
 for baryon non-conserving interactions such as
$e^{-} + d  \leftrightarrow {\bar u} + {\bar u}$,  etc.,
or decays of the supermassive
 gauge bosons X and Y such as
$ X  \rightarrow e^{-} + d, {\bar u} + {\bar u}$.
 Although today these interactions
are very ineffective because of the very large masses of the X
and Y bosons, in the early Universe when
$T \sim M_{ X} \sim 10^{15}$  GeV these types of interactions
should have been very important.
 C and CP violation is very model dependent.  In the minimal SU(5)
model, as we will see,
the magnitude of C and CP violation is too small to yield a useful value of
$\eta$.  The C and CP violation in general  comes
from the interference between
 tree level and first loop corrections.

The departure from equilibrium is very common in the
early Universe when interaction
rates cannot keep up with the expansion rate.  In fact,
the simplest (and most useful)
scenario for baryon production makes use of the fact that a
single decay rate goes out of equilibrium.  It is commonly referred to
 as the out of equilibrium decay scenario
\cite{ww}.  The basic idea is that the gauge bosons
 $X$ and $Y$ (or Higgs bosons)
 may have a lifetime long enough to insure that the
inverse decays have already
ceased so that the baryon number is produced by their free decays.

More specifically, let us call $X$, either the gauge
boson or Higgs boson, which produces
the baryon asymmetry through decays.  Let
$\alpha$  be its coupling to fermions.  For $X$ a gauge boson,  $\alpha$
will be the GUT fine structure constant, while for $X$ a Higgs boson,
$(4{\pi \alpha })^{ 1/2}$  will be the Yukawa coupling to fermions.
 The decay rate for $X$ will be
\beq
 \Gamma_{ D}  \simeq   \alpha M_{X}
\eeq
  However decays can only begin occurring when the age
of the Universe is longer
 than the $X$ lifetime
$\Gamma_D^{-1}$,  i.e., when  $\Gamma_{ D} > $ H
\beq
  \alpha M_{ X}  \ga  N(T)^{ 1/2} T^2/M_{ P}
\eeq
 or at a temperature
\beq
 T^{ 2}  \la  \alpha M_{ X}M_{ P}N(T)^{ -1/2}.
\eeq
Scatterings on the other hand proceed at a rate
$\Gamma_{ S}  \sim \alpha^2 T^3/M_X^2$
 and hence are not effective at lower temperatures.  To be in equilibrium,
decays must have been effective as T fell below
$M_{ X}$  in order to track the equilibrium
density of $X$'s (and  ${\bar X}$'s).
Therefore, the out-of-equilibrium condition  is
that at $T = M_{ X},   \Gamma {}_{ D} < H$
or
 \beq
M_{ X} \ga  \alpha M_{ P} (N(M_{ X}))^{ -1/2}
\sim 10^{18} \alpha {\rm GeV}
\label{mxmin}
\eeq
 In this case, we would expect a maximal net baryon asymmetry to be produced.

To see the role of C and CP violation, consider the
two channels for the decay
of an X gauge boson: $ X  \rightarrow (1) {\bar u}{\bar u}, (2) e^{-} d$.
Suppose that the branching ratio into the first channel with baryon number
$B = - 2/3$ is $r$ and that of the second channel with baryon number
$B = + 1/3$ is $1-r$. Suppose in addition that the branching ratio
for ${\bar X}$ into $({\bar 1})$ $u$ $u$ with baryon number
$B = + 2/3$ is ${\bar r}$ and
into $({\bar 2})$ $e^+ {\bar d}$ with baryon
number $B = - 1/3$ is $1- {\bar r}$.
Though the total decay rates of $X$ and ${\bar X}$ (normalized to unity)
are equal as required by CPT invariance, the differences in the individual
branching ratios signify a violation of C and CP conservation.

The (partial) decay rate for $X$ is computed from an invariant
transition rate
\beq
W = {s \over 2^n} |{\cal M}|^2 (2 \pi)^4 \delta^4 (\Sigma P)
\eeq
where the first term is the common symmetry factor
and the decay rate is
\beq
\Gamma = { 1 \over 2 M_X} \int W d\Pi_1 d\Pi_2
\eeq
with
\beq
d\Pi = {g d^3p \over (2\pi)^3 2 E}
\eeq
for $g$ degrees of freedom.
 Denote the parity (P)
of the states (1) and (2) by $\uparrow$ or
$\downarrow$, then we have the following transformation properties:
\beq
\begin{array}{lccc}
{\rm Under~CPT:} \qquad & \Gamma(X \rightarrow 1 \uparrow) & = &
\Gamma(  {\bar 1} \downarrow \rightarrow {\bar X}) \\
{\rm Under~CP:} \qquad & \Gamma(X \rightarrow 1 \uparrow) & = &
\Gamma({\bar X} \rightarrow {\bar 1} \downarrow) \\
{\rm Under~C:} \qquad & \Gamma(X \rightarrow 1 \uparrow) & = &
\Gamma( {\bar X} \rightarrow {\bar 1} \uparrow )
\end{array}
\label{sym}
\eeq
We can now denote
\begin{eqnarray}
r = \Gamma(X \rightarrow 1 \uparrow) + \Gamma(X \rightarrow 1 \downarrow) \\
{\bar r} = \Gamma( {\bar X} \rightarrow {\bar 1} \uparrow )
+ \Gamma({\bar X} \rightarrow {\bar 1} \downarrow)
\end{eqnarray}
The total baryon number produced by an $X$, ${\bar X}$ decay
is then
\begin{eqnarray}
\Delta B & = & -{2 \over 3} r + {1 \over 3} (1 - r) + {2 \over 3} {\bar r}
- {1 \over 3} (1 - {\bar r}) \nonumber \\
& = & {\bar r} - r = \Gamma( {\bar X} \rightarrow {\bar 1} \uparrow )
+ \Gamma({\bar X} \rightarrow {\bar 1} \downarrow) -
\Gamma(X \rightarrow 1 \uparrow) - \Gamma(X \rightarrow 1 \downarrow)
\end{eqnarray}
One sees clearly therefore, that from eqs. (\ref{sym}) if {\em either}
C {\em or} CP are good symmetries, $\Delta B = 0$.

In the out-of-equilibrium decay scenario \cite{ww}, the total baryon asymmetry
produced is proportional to $\Delta B = ({\bar r} - r)$.
If decays occur out-of-equilibrium, then at the time of decay,
$n_X \approx n_\gamma$ at $T < M_X$. We then have
\beq
{n_B \over s} = {(\Delta B) n_X \over s} \sim
{(\Delta B) n_X \over N(T) n_\gamma} \sim 10^{-2}(\Delta B)
\label{nbmax}
\eeq

The schematic view presented above can be extended to a complete
calculation given a specific model \cite{fot,kw}, see also
\cite{kt} for reviews. It makes sense to first consider the simplest
GUT, namely SU(5) (for a complete discussion of GUTs see \cite{rossbook}. In
SU(5), the standard model fermions are placed in a ${\bar 5}$ and $10$
representation of SU(5)
\beq
\left( \begin{array}{c}
d_1^c \\
d_2^c \\
d_3^c \\
e \\
\nu \end{array} \right)_L = {\bar {\bf 5}} \qquad
\left( \begin{array}{ccccc}
0 & u_3^c & -u_2^c & -u_1 & -d_1 \\
 & 0 & u_1^c & -u_2 & -d_2 \\
 & & 0 & -u_3 & -d_3 \\
 & & & 0 & -e^c \\
 & & & & 0 \end{array} \right)_L = {\bf 10}
\eeq
where the subscripts are SU(3)-color indices.  The standard model gauge
sector is augmented by the color triplet X and Y gauge bosons
which form a doublet under SU(2)$_L$ and have electric charges $\pm 4/3$
and $\pm 1/3$ respectively. The full set of 24 gauge bosons are in the adjoint
representation. In minimal SU(5), an adjoint of Higgs scalars, $\Sigma$,
is required for the breakdown of SU(5) to the standard model
SU(3)$_c \times$ SU(2)$_L \times$ U(1)$_Y$.  The additional Higgs scalars
needed to break the standard model down to SU(3)$_c \times$ U(1)$_{em}$
requires a five-plet of scalars, $H$, which contains the standard model
Higgs doublet in addition to a colored (charged $\pm 1/3$) triplet.

The SU(5) gauge couplings to fermions can be written as \cite{begn}
\begin{eqnarray}
{1 \over \sqrt{2}}g_5 {X_i}_\mu \left( {\bar {d_i}_R} \gamma^\mu e_R^+
+\epsilon_{ijk}{\bar {u_k^c}_L} \gamma^u {u_j}_L  +
{\bar {d_i}_L} \gamma^\mu e_L^+ \right) \\
{1 \over \sqrt{2}}g_5 {Y_i}_\mu \left( -{\bar {d_i}_R} \gamma^\mu {\nu^c}_R
+\epsilon_{ijk}{\bar {u_j^c}_L} \gamma^u {d_k}_L  -
{\bar {u_i}_L} \gamma^\mu e_L^+ \right)
\end{eqnarray}
where $g_5$ is the SU(5) gauge coupling constant.
These couplings lead to the decays shown in figure 5. Similar diagrams
can be drawn for the decay of the Y gauge boson.

\begin{figure}
\vspace{6cm}
\caption{Decay diagrams for the X gauge boson.}
\end{figure}

The Higgs five-plet, $H$ couples to fermions via the
\beq
{\bf H~{\bar 5}~10} \qquad {\bf H~10~10}
\eeq
couplings shown in figure 6 (shown are the
couplings of the triplet relevant for
baryogenesis).

\begin{figure}
\vspace{6cm}
\caption{Higgs couplings to fermions.}
\end{figure}

Typically, it is expected that the Higgs masses, in particular, those of the
adjoint, $\Sigma$, are of order the GUT scale, $M_X \sim 10^{15} - 10^{16}$
GeV. The five-plet is somewhat problematic however, as the the doublet in $H$,
must remain light as it corresponds to the standard model electroweak Higgs
doublet.  The triplet can not be light because as a consequence of
the diagrams in figure 6, it will mediate proton decay.  However,
because the couplings to fermions in figure 6 are Yukawa couplings rather than
gauge couplings, the calculated rate for proton decay mediated by the triplets
will be much smaller, allowing for a smaller triplet mass
\beq
{\Gamma(p-{\rm decay via} X) \over \Gamma(p-{\rm decay via} H)}
\sim \left({M_H \over M_X}\right)^4 \left({M_W \over m_q} \right)^4
\label{lowmassh}
\eeq
implying that the Higgs triplet mass $M_H$ need only be greater
than about $10^{10}$ GeV.

{}From equation (\ref{nbmax}) it is clear that a complete calculation of
$n_B/s$ will require a calculation of the CP violation in the decays
(summed over parities) which we can parameterize by
\beq
\epsilon = {\bar r} - r = {\Gamma({\bar X} \rightarrow {\bar 1})
- \Gamma(X \rightarrow 1) \over \Gamma({\bar X} \rightarrow {\bar 1})
+ \Gamma(X \rightarrow 1)} \sim {{\rm Im} \Gamma \over {\rm Re} \Gamma}
\eeq
At the tree level, as one can see $\Gamma(X \rightarrow 1) \propto
g_5^\dagger g_5$ is real and there is no C or CP violation.
At the one loop level, one finds that the interference between
the tree diagram and the loop diagram shown in figure 7 gives \cite{nw}
\beq
\epsilon \propto {\rm Im} g_{X_1}^\dagger g_{Y_1} g_{X_2} g_{Y_2}^\dagger
\eeq
However in SU(5), $g_{X_1} = g_{Y_1} = g_{X_2} = g_{Y_2} = g_5$
so that
\beq
\epsilon \propto {\rm Im} (g_5^\dagger g_5) (g_5 g_5^\dagger) = 0
\eeq
Similarly, the exchange of the Higgs triplet at one loop also
gives a vanishing contribution to $\epsilon$.

\begin{figure}
\vspace{6cm}
\caption{One loop contribution to the C and CP violation in SU(5).}
\end{figure}

At least two Higgs five-plets are therefore required to generate sufficient
C and CP violation. (It is possible within minimal SU(5) to generate a non-
vanishing $\epsilon$ at 3 loops, however its magnitude would be too small
for the purpose of generating a baryon asymmetry.)  With two five-plets,
$H$ and $H^\prime$, the interference of diagrams of the type in figure 8,
will yield a non-vanishing $\epsilon$,
\beq
\epsilon \propto {\rm Im} ({a^\prime}^\dagger a b^\prime b^\dagger) \ne 0
\eeq
if the couplings $a \ne a^\prime$ and $b \ne b^\prime$.

\begin{figure}
\vspace{6cm}
\caption{One loop contribution to the C and CP
violation with two Higgs five-plets.}
\end{figure}

Given the grand unified theory, SU(5) in this case, the final task in
computing the baryon asymmetry is to take into account the thermal
history of the Universe and the departure from thermal equilibrium
\cite{yosh,ww}. A full complete numerical calculation was undertaken in
\cite{fot} and these results will be briefly summarized here.

To trace the evolution of the baryon asymmetry contained in quarks,
a full set of coupled differential Boltzmann equations must be
computed for all relevant particle species.  In general, particle number
densities must satisfy
\begin{eqnarray}
\dot{n} + 3 H n & = & \int d\Pi_a d\Pi_b \cdots
 d\Pi_i d\Pi_j \cdots  \nonumber \\
& & \left[ f_a f_b \cdots (1 \pm f_i) (1 \pm f_j) \cdots
W(p_a p_b  \cdots \rightarrow p_ip_j \cdots ) \right. \nonumber \\
& & \left. -f_i f_j \cdots (1 \pm f_a) (1 \pm f_b) \cdots
W(p_i p_j  \cdots \rightarrow p_ap_b \cdots ) \right]
\end{eqnarray}
where
\beq
n = 2 \int E f d\Pi \qquad f = {F(t) \over e^{E/T} \pm 1}
\eeq
is the number density of particles and the energy distribution.
 In thermal equilibrium $F = 1$
and is allowed to take other values. Since we are interested in an asymmetry
it is more convenient to keep track of the quantities
\beq
n_{i+} = n_i + n_{\bar i} \qquad n_{i-} = n_i - n_{\bar i}
\eeq
For small asymmetries, $F_+ \simeq 2$ and $F_-$ is small.
In total, it is necessary to keep track of the following 12
quantities: $U_+,D_+, L_+, \nu_+, X_+, Y_+; U_-,D_-, L_-, \nu_-, X_-, Y_-$
where these scaled functions are defined by $U(t) = n_u/(g_u A)$ with
$A= [3\zeta(3)/4\pi^2]T^3$.
It is also convenient to change time variables from ${d \over dt}$ to
${d \over dz} = {5.8 \times 10^{17} {\rm GeV} \over M_X^2} \left({160
\over N(T)} \right)^{1/2} z {d \over dt}$, with $z = M_X/T$.

The full set of coupled equations can be found in \cite{fot}.
For our purposes here, it will be useful to write down only a sample
equation for $U_-$
\begin{eqnarray}
 {1 \over zK}{dU_- \over dz}  & = & - \gamma_{\rm D} \left[
2X_- +  Y_-/2 \right] - \gamma_{\rm ID} \left[ 2U_+ U_-
+ (U_+D_- + D_+U_-)/2 \right. \nonumber \\
& & \left. (U_+L_- + L_+U_-)/4 \right]
 +\epsilon \left[ \gamma_{\rm D} X_+ - \gamma_{\rm ID} D_+ L_+ /2
\right] + {\rm scatterings}
\label{u-}
\end{eqnarray}
where
\begin{eqnarray}
\gamma_{\rm D} = {1 \over \alpha M_X} {4 \over 3}
{ \int d \Pi_X d\Pi_{u_1} d \Pi_{u_2} f_X W(X \rightarrow u_1 u_2)
\over g_X \zeta(3) T^3/\pi^2} \\
\gamma_{\rm ID} = {1 \over \alpha M_X} {4 \over 3}
{ \int d \Pi_X d\Pi_{u_1} d \Pi_{u_2} f_{u_1}
f_{u_2} W(u_1 u_2 \rightarrow X)
\over g_X \zeta(3) T^3/\pi^2} \\
K = {2.9 \times 10^{17} \alpha {\rm GeV} \over M_X} \left({160
\over N(T)} \right)^{1/2}
\end{eqnarray}
where the gut fine-structure constant is $\alpha = g_5^2/4 \pi$.

When equilibrium is maintained and all interaction rates are
large compared with the expansion rate, solutions to the $+$ equations
(not shown) give $U_+, D_+, L_+, \nu_+ = 2$ and
$X_+ = Y_+ = 2 \gamma_{\rm ID}/\gamma_{\rm D}$. In
this case, as one can see from the
sample $-$ equation in (\ref{u-}), all CP violation effects disappear
(the coefficient of $\epsilon$ vanishes). As the $\epsilon$ term was the
only one that could generate an asymmetry, the asymmetry is driven
to 0 in equilibrium.

To get a feeling for the results of such a numerical integration, let
us first consider the case with $\epsilon = 0$. When $ B - L = 0$ initially,
there is a damping of any initial baryon asymmetry as is shown in figure 9.
The parameter $z$, increases as a function of time ($z \sim \sqrt{t}$).
In equilibrium, the asymmetries are damped until the baryon number violating
interactions freeze-out. In accord with our earlier remarks, a large
value of $M_X$ (corresponding to a small value of $K$) results in an early
departure from equilibrium and a larger final baryon asymmetry. If $B-L
\ne 0 $ initially, since the minimal SU(5) considered here conserves $B-L$,
the asymmetry can not be erased, only reshuffled.

\begin{figure}
\vspace{11.5cm}
\caption{The damping of an initial baryon asymmetry
with $B-L=0$ and $\epsilon
=0$.}
\end{figure}

To generate an asymmetry, we must have $\epsilon \ne 0$. The time evolution
for the generation of a baryon asymmetry is shown in figure 10.  As one
can see, for large values of $M_X$, ie. values which satisfy the lower
limit given in eq. (\ref{mxmin}), the maximal value for the baryon asymmetry,
$n_B/s \sim 10^{-2} \epsilon$ is achieved.  This confirms numerically
the original out-of equilibrium decay scenario \cite{ww}.  For smaller
values of $M_X$, an asymmetry is still produced, which
however is smaller due to  partial equilibrium
maintained by inverse decays ($\gamma_{\rm ID}$).  The growth of the
asymmetry
as a function of time is now damped, and it reaches its final value when
inverse decays freeze out.

\begin{figure}
\vspace{11cm}
\caption{The time evolution of the baryon asymmetry
with $B = L = 0$ initially.}
\end{figure}

Finally it is important to note that the results for the final baryon
asymmetry, as shown in figure 10, as a function of the mass of the $X$
gauge boson, is in fact largely independent of the initial baryon asymmetry.
This is evidenced in figure 11, which shows the time
evolution of the asymmetry,
given a large initial asymmetry.  Even for
large $M_X$, the asymmetry is slightly
damped, and for smaller values of $M_X$, the asymmetry is damped to a level
which is again determined by the freeze-out of inverse decays.
This means that this mechanism of baryogenesis is truely
independent of initial
conditions, in particular it gives the same value for
$\eta$ whether or not $\eta = 0$ or 1 initially.

\begin{figure}
\vspace{11cm}
\caption{The time evolution of the baryon asymmetry with a large initial
baryon asymmetry.}
\end{figure}

The out-of-equilibrium decay scenario discussed above did not include
the effects of an inflationary epoch.
In the context of inflation,
 one must in addition ensure baryogenesis
after inflation as any asymmetry produced before inflation
would be inflated away along with magnetic monopoles and any other unwanted
relic.
 Reheating after inflation, may require
 a Higgs sector with a relatively light $O(10^{10}-10^{11}) GeV$ Higgs boson.
The light Higgs
is necessary since the inflaton, $\eta$, is typically very light ($m_\eta
\sim \mu^2/M_P \sim
O(10^{11})$ GeV, determined from the magnitude of density
perturbations on large scales as measured by COBE\cite{cobe},
 cf. eq. (\ref{cobemu}))
and the baryon number violating Higgs
would have to be produced during inflaton decay.
Note that a ``light" Higgs is acceptable as discussed above due to the
reduced couplings to fermions cf. eq.(\ref{lowmassh}).
The out-of-equilibrium decay scenario would now be realized by
Higgs boson decay rather than gauge boson decay and a different sequence
of events. First the inflaton would be required to
decay to Higgs bosons (triplets?) and subsequently
the triplets would decay rapidly by the processes shown in figure 6.
These decays would be well out of equilibrium
as at reheating $T \ll m_H$ and $n_H \sim n_\gamma$ \cite{dlnos}.
In this case, the baryon asymmetry is given simply by
\beq
{n_B \over s} \sim \epsilon {n_H \over {T_R}^3}
\sim \epsilon {n_\eta \over {T_R}^3}
\sim \epsilon {T_R \over m_\eta} \sim \epsilon
\left( {m_\eta \over M_P} \right)^{1/2}
\sim \epsilon {\mu \over M_P}\sim 10^{-4} \epsilon
\eeq
where $T_R$ is the reheat temperature after inflation,
and I have substituted for $n_\eta = \rho_\eta/m_\eta
\sim \Gamma^2{M_P}^2/m_\eta$.

\subsection{Supersymmetry}

Supersymmetry, as is well known by now, was incorporated into GUTs because
of its ability to resolve the gauge hierarchy problems.  There are two
aspects to this problem:  1) there is a separation in physical mass scales,
$M_W  \ll M_X < M_P$; 2) this separation is extremely sensitive to radiative
corrections.  The first problem has to do with a tree-level choice
of mass parameters.  A single fine-tuning.  The second problem requires
fine-tuning at many successive orders in perturbation theory.
Radiative corrections to scalar masses are quadratically
divergent
\beq
	\delta m_o^2  \sim g^2 \int {d^4k \over (2 \pi^4)} {1 \over k^2 }
\sim 0(\alpha/\pi) \Lambda^2
\eeq
where $\Lambda$ is some cut-off scale.  In the low energy electroweak theory,
the
smallness of $M_W$  requires the mass of the physical
Higgs boson to be $m_H <
0(1)$TeV.  Requiring $\delta m_H^2  < 0(m_H^2)$ implies that
$\Lambda < 0(1)$ TeV as well.  The
trouble comes when we move to a GUT where the natural
 cut-off is $M_X$  (or even
$M_P$ ) rather than $0(M_W)$ and we expect $\delta m_H^2  > 0(10^{15})$
GeV.  A
cancellation may be imposed by hand, but this must be done to each order in
perturbation theory.  A solution to this difficulty would be to cancel the
radiative corrections by including fermion loops which have the opposite
sign.  Then provided $|m_B^2  - m_F^2| < 0(1)$ TeV, the stability of
the mass scales would be guaranteed.  Such a cancellation occurs
automatically in a supersymmetric theory (in the limit of exact
supersymmetry, these radiative corrections are absent entirely).  In
addition, although gauge couplings still get renormalized, the Yukawa
couplings of theory, which are parameters of a superpotential do not get
renormalized\cite{noren}.

Standard unification (ie. non-supersymmetric) has come across additional
difficulties of late. Extrapolation of the gauge
coupling constants of the standard
model using the renormalization group equation with standard model inputs,
does not result in the three couplings meeting at a single point.
However, when the superpartners of the standard model fields are also
incorporated, and the renormalization group equations are again run to high
energy scales, then the  gauge couplings do in fact meet at a
point (within errors) at a scale of order $10^{16}$ GeV \cite{amaldi}.

The field content of the supersymmetric standard model, consists
of the following chiral supermultiplets: $Q, u^c, d^c, L, e^c,
H_1, H_2$.  The only addition is the extra Higgs doublet.
The Yukawa interactions are generated by the superpotential
\beq
F_Y = h_u H_1 Q u_c + h_d H_2 Q d^c + h_l H_2 L e^c + \epsilon H_1 H_2
\eeq
leading to the Lagrangian interactions
\beq
{\cal L} \ni \left( \partial^2F_Y / \partial \phi^i \partial \phi^j \right)
\Psi^i \Psi^j
\label{yuk}
\eeq
where $\Psi^i$  is the fermion component of the  superfield $\phi^i$.
(\ref{yuk}) contains the normal fermion mass terms of the standard model.

	The scalar potential in a globally supersymmetric theory can be written
as
\beq
	V(\phi^i,\phi_i^*) = \Sigma_i |F_i|^2  +   {1 \over 2}
\Sigma_a g_a^2 |D^a|^2
\label{scalpot}
\eeq
where
\beq
	F_i =  \partial F/ \partial \phi^i
\eeq
for superpotential $F$ and
\beq
	|D^a|^2  = \left(\Sigma_{i,j}  \phi_i^* {T^a}_j^i \phi^j \right)^2
\eeq
for generators ${T^a}_j^i$ of a gauge group with gauge coupling $g_a$.
In addition, in broken supersymmetry there will be
 soft supersymmetry breaking
scalar masses as well as gaugino masses.

In a supersymmetric grand unified SU(5)
 theory, the superpotential $F_Y$ can be expressed in terms
of SU(5) multiplets
\beq
	F_Y  = h_d {\bf H_2 ~{\bar 5}~10}  +  h_u {\bf H_1~10~10}
\eeq
where $10, {\bar 5}, H_1$ and $H_2$ are chiral
supermultiplets for the 10, and ${\bar 5}$ plets of
SU(5) matter fields and the Higgs
5 and ${\bar 5}$ multiplets respectively.

In supersymmetric SU(5), there are now new dimension 5
operators which violate baryon number and lead to proton decay
as shown in figure 12.
The first of these diagrams leads to effective dimension 5 Lagrangian terms
such as
\beq
 {\cal L}_{\rm eff}^{(5)} = {h_u h_d \over M_{H}} ( \tilde q
\tilde q q l)
\eeq
and the resulting dimension 6 operator for proton decay \cite{enr}
\beq
{\cal L}_{\rm eff} = {h_u h_d \over M_{H}} \left( {g_2^2 \over M_{\tilde W}}
 {\rm or} {g_1^2 \over M_{\tilde B}} \right) (  q q q l)
\eeq
As a result of these diagrams the proton decay rate scales as $\Gamma
\sim h^4 g^4/M_H^2 M_{\tilde G}^2$ where $M_H$ is the triplet mass, and
$M_{\tilde G}$ is a typical gaugino mass of order $\la$ 1 TeV.  This rate
however is much too large if $M_H \sim 10^{10}$ GeV.

\begin{figure}
\vspace{6cm}
\caption{Dimension 5 and induced dimension 6 graphs violating baryon number.}
\end{figure}

It is however possible to have a lighter ($O(10^{10}-10^{11})$ GeV)
Higgs triplet needed for baryogenesis in the out-of-equilibrium decay
scenario with inflation.  One needs  two pairs of Higgs five-plets
($H_1, H_2$ and  $H_1^\prime, H_2^\prime$ which
is anyway necessary to have sufficient
C and CP violation in the decays.
By coupling one pair $(H_2$ and $H_1^\prime)$
only to the third generation of fermions
via \cite{nt}
\beq
a {\bf H_1 10 10} + b {\bf H_1^\prime 10_3 10_3} + c {\bf H_2 10_3 {\bar 5}_3}
+ d {\bf H_2^\prime 10 {\bar 5}}
\eeq
proton decay can not be induced by the dimension five operators.
Triplet decay will however generate a baryon asymmetry proportional to
$\epsilon \sim {\rm Im} d c^\dagger b a^\dagger$.

\subsection{The Affleck-Dine Mechanism}

Another mechanism for generating the  cosmological baryon asymmetry
is the decay of scalar condensates as first
proposed by Affleck and Dine\cite{21}.
This mechanism is truly a product of supersymmetry.
It is straightforward though tedious to show that
  there are many directions in field space such that the scalar potential
given in eq. (\ref{scalpot}) vanishes identically
 when SUSY is unbroken. That is, with a particular
assignment of scalar vacuum expectation values, $V=0$ in both the
$F-$ and $D-$ terms.  An example of such a direction
is
\beq
u_3^c = a \qquad s_2^c = a \qquad
-u_1 = v \qquad \mu^- = v \qquad b_1^c = e^{i\phi} \sqrt{v^2 + a^2}
\label{flat}
\eeq
where $a,v$ are arbitrary complex vacuum expectation values.
 SUSY breaking lifts this degeneracy so that
\begin{equation}
	V  \simeq \tilde{m}^2 \phi^2
\eeq
where $\tilde{m}$ is the SUSY breaking scale and $\phi$ is the direction
 in field space corresponding to the flat direction.
  For large initial values of $\phi$, \ $\phi_o \sim M_{gut}$,
 a large baryon asymmetry can be generated\cite{21,22}. This requires
the presence of baryon number violating operators such as $O=qqql$ such that
$\langle O \rangle \neq 0$.  The decay of these
 condensates through such an operator
can lead to a net baryon asymmetry.

In a supersymmetric gut, as we have seen above, there are
precisely these types of operators. In figure 13, a 4-scalar diagram involving
the fields of the flat direction (\ref{flat}) is shown. Again, $\tilde G$ is a
(light) gaugino. The two supersymmetry breaking insertions are of order
$\tilde m$, so that the diagram produces an effective quartic coupling
of order ${\tilde m}^2/(\phi_o^2 + M_X^2)$.

\begin{figure}
\vspace{6cm}
\caption{Baryon number violating diagram involving flat direction fields.}
\end{figure}

The baryon asymmetry produced, is computed by tracking the evolution of the
sfermion condensate, which is determined by
\begin{equation}
\ddot{\phi} + 3H\dot{\phi} = - {\tilde m}^2 \phi
\end{equation}
To see how this works, it is instructive to consider
a toy model with potential
\cite{22}
\beq
V(\phi,\phi^*)  = \tilde{m}^2 \phi \phi^* + {1 \over 2} i \lambda [\phi^4
-{\phi^*}^4 ]
\label{toy}
\eeq
The equation of motion becomes
\begin{eqnarray}
\ddot{\phi_1} + 3H\dot{\phi_1} = - {\tilde m}^2 \phi_1 +
3 \lambda \phi_1^2 \phi_2
-\lambda \phi_2^3 \\
\ddot{\phi_2} + 3H\dot{\phi_2} = - {\tilde m}^2 \phi_2 -
3 \lambda \phi_2^2 \phi_1
+ \lambda \phi_1^3
\end{eqnarray}
with $\phi = (\phi_1 + i \phi_2)/\sqrt{2}$.
Initially, when the expansion rate of the Universe, $H$, is large, we
can neglect $\ddot \phi$ and $\tilde m$. As one can see from (\ref{toy})
the flat direction lies along $\phi \simeq \phi_1 \simeq \phi_o$
with $\phi_2 \simeq 0$. In this case, $\dot{\phi_1} \simeq 0$ and
$\dot{\phi_2} \simeq {\lambda \over 3H} \phi_o^3$.  Since the baryon density
can be written as $n_B = j_o = {1 \over 2} ( \phi_1 \dot{\phi_2} -
\phi_2 \dot{\phi_1} ) \simeq {\lambda \over 6H} \phi_o^4$, by generating
some motion in the imaginary $\phi$ direction, we have generated a net baryon
density.

When $H$ has fallen to order $\tilde m$ (when $t^{-1} \sim \tilde m$),
$\phi_1$ begins to oscillate about the origin with $\phi_1 \simeq
\phi_o \sin (\tilde {m} t)/\tilde{m}t$ At this
 point the baryon number generated is
conserved and the baryon density, $n_B$ falls as $R^{-3}$.  Thus,
\beq
n_B \sim {\lambda \over \tilde m} \phi_o^2  \phi^2 \propto R^{-3}
\eeq
and relative to the number density of $\phi$'s
($n_\phi = \rho_\phi / \tilde m
= \tilde{m} \phi^2$)
\beq
{n_B \over n_\phi} \simeq {\lambda \phi_o^2 \over {\tilde m}^2}
\eeq

If it is assumed that the energy density of the Universe is dominated
by $\phi$, then the oscillations will cease, when
\beq
\Gamma_\phi \simeq {{\tilde m}^3 \over \phi^2} \simeq H
\simeq {\rho_\phi^{1/2} \over M_P} \simeq {{\tilde m} \phi \over M_P}
\eeq
or when the amplitude of oscillations has dropped to $\phi_D \simeq (M_P
{\tilde m}^2 )^{1/3}$. Note that the decay rate is suppressed as
fields coupled directly to $\phi$ gain masses $\propto \phi$.
It is now straightforward to compute the baryon to entropy ratio,
\beq
{n_B \over s} = {n_B \over \rho_\phi^{3/4}} \simeq {\lambda \phi_o^2
\phi_D^2 \over {\tilde m}^{5/2} \phi_D^{3/2}} =
{\lambda \phi_o^2 \over {\tilde m}^2} \left({M_P \over \tilde m}\right)^{1/6}
\eeq
and after inserting the quartic coupling
\beq
{n_B \over s} \simeq \epsilon {\phi_o^2 \over (M_X^2 + \phi_o^2)}
\left({M_P \over \tilde m}\right)^{1/6}
\eeq
which could be quite large.

In the context of inflation, a couple of significant changes to the scenario
take place. First, it is more likely that the energy density
is dominated by the inflaton rather than the sfermion condensate.
Second, the the initial value (after inflation) of the condensate
$\phi$  can be determined by the inflaton mass $m_\eta$,
${\phi_o}^2 \simeq H^3\tau \simeq m_\eta M_P$.
The sequence of events leading to a baryon
 asymmetry is then as follows \cite{eeno}:
After inflation, oscillations of of the inflaton begin at $R=R_\eta$
when $H \sim m_\eta$ and oscillations of the sfermions begin at
$R=R_\phi$ when $H\sim {\tilde m}$. If the Universe is inflaton dominated,
$H \sim m_\eta (R_\eta / R)^{3/2}$ since $H \sim \rho_\eta^{1/2}$ and
$\rho_\eta \sim \eta^2 \sim R^{-3}$ Thus one can relate $R_\eta$ and $R_\phi$,
$R_\phi \simeq (m_\eta / {\tilde m})^{2/3} R_\eta$. As discussed earlier,
inflatons decay when $\Gamma_\eta = m_\eta^3/M_P^2 = H$ or when
$R=R_{d\eta} \simeq (M_p/m_\eta)^{4/3} R_\eta$.
The Universe then becomes dominated
by the relativistic decay products of the inflaton,
$\rho_{r\eta} = m_\eta^{2/3} M_P^{10/3} (R_\eta/R)^4$ and
$H = m_\eta^{1/3} M_P^{2/3} (R_\eta/R)^2$.
Sfermion decays still occur when $\Gamma_\phi =H$ which now
corresponds to a value of the scale factor $R_{d\phi}
=(m_\eta^{7/15} \phi_o^{2/5} M_P^{2/15}/{\tilde m}) R_\eta$.

  Finally, the baryon asymmetry in the Affleck-Dine
 scenario with inflation becomes \cite{eeno}
\begin{equation}
	\frac{n_B}{s} \sim  \frac{\epsilon {\phi_o}^4 {m_\eta}^{3/2}}
{{M_X}^2 {M_P}^{5/2} \tilde{m}} \sim
 \frac{\epsilon m_\eta^{7/2}}{{M_X}^2 {M_P}^{1/2} \tilde{m}}
  \sim  (10^{-6}-1) \epsilon
\end{equation}
for
 $\tilde{m} \sim (10^{-17}-10^{-16}) M_P$,
 and $M_X \sim (10^{-4}-10^{-3}) M_P$
and $m_\eta \sim (10^{-8} - 10^{-7} ) M_P$.

\section{Lepto-Baryogenesis}

The realization\cite{6} of significant baryon number
violation at high temperature within the
standard model, has opened the door for many new
possibilities for the generation
of a net baryon asymmetry. Indeed, it may be possible
to generate the asymmetry
entirely with the context of the standard model \cite{ckn}. Electroweak
 baryon number violation
occurs through non-perturbative interactions mediated by ``sphalerons",
 which violate $B + L$ and conserve
$B - L$.  For this reason, any gut produced asymmetry with $B - L = 0$
may be subsequently erased by sphaleron interactions \cite{14}.

The origin of the sphaleron interactions lies in the anomalies
of the electroweak current
\beq
J_B^\mu = N_f \left( {g_2^2 \over 32 \pi^2} W \tilde{W} -
{g_1^2 \over 32 \pi^2} B \tilde{B} \right)
\eeq
This gives rise to a non-trivial vacuum structure with degenerate
vacuum states with differing baryon number. At $T=0$,
the rates for such transitions is highly suppressed \cite{thoo},
$\propto e^{-2\pi /\alpha_W}$.  However at high temperatures,
the transition rate is related to the diffusion rate over a potential
barrier, $\Gamma_S \sim \left(M_W^7/\alpha_W^3T^6 \right) e^{-4M_W/
\alpha_W T}$ in the broken phase.
In the symmetric phase, the barrier becomes very small and transitions
are relatively unsuppressed, $\Gamma_S \sim \left( \alpha_W^4 T \right)$.

With $B- L = 0$, it is relatively straightforward to see
that the equilibrium conditions
including sphaleron interactions gives zero net baryon number \cite{27}.
By assigning each particle species a chemical potential, and
using gauge and Higgs
interactions as conditions on these potentials
(with generation indices suppressed),
\begin{eqnarray}
\mu_- + \mu_0 = \mu_W \qquad
\mu_{u_R} - \mu_{u_L} = \mu_0 \qquad
\mu_{d_R} - \mu_{d_L} = -\mu_0 \nonumber \\
\mu_{l_R}-\mu_{l_L} = -\mu_0 \qquad
\mu_{d_L} - \mu_{u_L} = \mu_W \qquad
\mu_{l_L} - \mu_\nu =  \mu_W
\end{eqnarray}
one can write
down a simple set of equations for the baryon and
lepton numbers and electric charge which reduce to:
\begin{eqnarray}
B& = & 12 \mu_{u_L} \nonumber \\
L& = & 3 \mu - 3 \mu_0 \label{mus} \\
Q& = & 6 \mu_{u_L} -2\mu + 14 \mu_0 \nonumber
\end{eqnarray}
where $\mu = \sum \mu_{\nu_i}$.
In (\ref{mus}), the constraint on the weak isospin charge,
$Q_3 \propto \mu_W = 0$ has been employed.
Though the charges $B,L,$ and $Q$ have been written as
chemical potentials, since for small asymmetries,
an asymmetry $(n_f - n_{\bar f})/s \propto \mu_f/T$,
we can regard these quantities as net number densities.

The sphaleron process yields the additional condition,
\beq
 9\mu_{u_L} + \mu = 0
\label{S}
\eeq
which allows one to solve for $L$ and $B-L$ in terms of
$\mu_{u_L}$, ultimately giving
\beq
B = {28 \over 79} \left( B - L \right)
\label{2879}
\eeq
Thus, in the absence of a primordial $B-L$ asymmetry,
the baryon number is erased by equilibrium processes.
Note that barring new interactions (in an extended model)
the quantities ${1 \over 3}B - L_e$, ${1 \over 3}B - L_\mu$,
and ${1 \over 3}B - L_\tau$ remain conserved.

With the possible erasure of the baryon asymmetry when $B-L=0$ in mind,
since minimal SU(5) preserves $B-L$, electroweak
effects require guts beyond
SU(5) for the asymmetry generated by the out-of-equilibrium
decay scenario to survive.
Guts such as SO(10) where a primordial $\bl$ asymmetry
can be generated becomes a promising choice.
The same holds true in the Affleck-Dine mechanism for generating a
baryon asymmetry.
  In larger guts there are baryon number violating operators
and associated flat directions\cite{24}.  A specific
example in SO(10) was worked out in detail by Morgan\cite{25}.

An important question remaining to be answered is whether or not the
baryon asymmetry can in fact be generated during the electroweak
weak phase transition.  This has been the focus of much attention
in recent years. I refer the reader to the review of ref. \cite{ckn}.
In the remainder of these lectures, I will focus on alternative
means for generating a baryon asymmetry which none-the-less makes
use of the sphaleron interactions.

The above argument regarding the erasure of a primordial baryon asymmetry
 relied on the assumption that all particle species are
in equilibrium.  However, because of the extreme smallness of the electron
Yukawa coupling, $e_R$ does not come into equilibrium until the late times.
The $e_R$ decoupling temperature is determined by
the rate of $e_R \rightarrow e_L + H$ transitions
and comparing this rate to the expansion rate
\beq
\Gamma_{LR} = {\pi h_e^2 \over 192 \zeta(3)} {m_H^2 \over T} \sim
{20 T^2 \over M_P} \simeq H
\eeq
which gives $T = T_* \sim$ O(few) TeV.  Thus one may ask
the question, whether or not
the baryon asymmetry may be stored in a primordial
$e_R$ asymmetry \cite{cdeo3}.
Because sphalerons preserve $B-L$, any lepton number stuck
in $e_R$ is accompanied
by an equal baryon number.
However, at temperatures
below the $e_R$ decoupling
temperature, baryon number will begin to be
destroyed so long as sphalerons are in equilibrium.
Sphalerons are in equilibrium from about the electroweak phase transition to
$T \sim 10^{12}$ GeV \cite{6}.  As it turns out, the $e_R$ (baryon) asymmetry
is exponentially sensitive to parameters of the model.

To clearly see the role of $e_R$ decoupling, it is helpful to look again
at the equations relating chemical potentials.
Above the scale $T_*$, the relation $\mu_{e_R} = \mu_{e_L} - \mu_0$
does not hold.  Instead there is an equilibrium solution \cite{clkao3}
\begin{eqnarray}
\mu_0 = {5 \over 153} ( \mu_{e_R} - \mu_{e_L} ) \nonumber \\
B = 12\mu_{u_L} = {44 \over 153} ( \mu_{e_R} - \mu_{e_L} )
\end{eqnarray}
One can quickly see now that below $T_*$, when
$\mu_{e_R} = \mu_{e_L} - \mu_0$ is respected, the only solution yields
$\mu_0 = B = L = 0$.
In terms of conserved quantities, above, $T_*$ we can write the
equilibrium solution for $B$ \cite{clkao3}
\beq
B_{eq} = {66 \over 481} \left( 3L_{e_R} + ({1 \over 3}B - L_e) \right)
\eeq
where $L_{e_R}$ is the lepton asymmetry stored in $e_R$'s and
$L_e$ is the total lepton asymmetry. Note
that this is independent of the initial baryon asymmetry.
Below $T_*$, the baryon asymmetry drops off exponentially
\beq
B = B_{eq} e^{-\int_{t_*}^{t_c} {711 \over 481} \Gamma_{LR} dt}
\eeq
integrated from $T_*$ to the electroweak phase transition where
sphaleron interactions quickly freeze out.
 With standard model
parameters, the baryon asymmetry is not preserved \cite{cdeo3,clkao3}.

Another possibility for preserving a primordial baryon asymmetry
when $\bl = 0$ comes if the asymmetry produced by scalar condensates
in the Affleck-Dine mechanism is large ($n_B/s \ga 10^{-2}$) \cite{dmo}.
After the decay of the A-D condensate, the baryon number is
shared among fermion and boson superpartners. However, in equilibrium,
there is a maximum chemical potential $\mu_f = \mu_B = \tilde m$
and  for a large asymmetry, the baryon number density stored in
fermions, $n_{B_f} = {g_f \over 6} \mu_f T^2$ is much less than the total
baryon density. The bulk of the baryon asymmetry is driven into the
$p=0$ bosonic modes  and a Bose-Einstein condensate is formed \cite{dkiri}.
The critical temperature for the formation of this condensate
is given by $n_B \simeq n_{B_b} + n_{B_c} = {g_b \over 3} \tilde{m} T_c^2$
so that,
\beq
n_{B_c} = {g_b \over 3} \left( 1 - \left( {T \over T_c} \right)^2 \right)
 T_c^2
\eeq
At $T < T_c$, most of the baryon number remains in a condensate and
for large $n_B$, the condensate persists down to temperatures
of order 100 GeV. Thus sphaleron interactions are shut off and
a primordial baryon asymmetry is maintained even with $\bl = 0$.
One should note however that additional sources of entropy are required
to bring $\eta$ down to acceptable levels.

As alluded to above, sphaleron interactions also allow
 for new mechanisms to produce a baryon asymmetry.
The simplest of such mechanisms
is based on the decay of a right handed neutrino-like state\cite{20}.
This mechanism
is certainly novel in that does not require grand unification at all.
By simply adding to the Lagrangian a Dirac and Majorana mass term
 for a new right handed neutrino state,
\beq
{\cal L} \ni M\nu^c\nu^c + \lambda H L \nu^c
\eeq
the out-of-equilibrium decays $\nu^c \rightarrow L +  H^*$
 and  $\nu^c \rightarrow L^* + H$ will generate a non-zero
lepton number $L \neq 0$. The out-out-equilibrium condition
for these decays translates to $10^{-3} \lambda^2 M_P < M$
and $M$ could be as low as $O(10)$ TeV.
(Note that once again in order to
have a non-vanishing contribution to the C and CP violation
in this process at 1-loop, at least 2 flavors of $\nu^c$ are required.
For the generation of masses of all three neutrino flavors,
3 flavors of $\nu^c$ are required.)
 Sphaleron effects can transfer this lepton asymmetry into a baryon
 asymmetry since now $B - L \neq 0$. A supersymmetric version of
this scenario
has also been described \cite{cdo,mur}.

	The survival of the asymmetry, of course depends on
 whether or not electroweak sphalerons can wash away the asymmetry.
 The persistence of lepton number violating interactions in conjunction
 with electroweak sphaleron effects could wipe out \cite{26}
 both the baryon and lepton asymmetry in the
 mechanism described above through effective operators
 of the form $\lambda^2 LLHH/M$.
In terms of chemical potentials, this interaction adds the condition
$\mu_\nu + \mu_0 = 0$.  The constraint comes about by requiring that
this interaction
be out of equilibrium at the time when sphalerons are in equilibrium.
 The additional
condition on the chemical potentials would force the solution $B=L=0$.

It is straightforward to derive a constraint
\cite{26}-\cite{29},\cite{27} on $ M/\lambda^2$. So long as the
$\Delta L = 2$ operator is out-of-equilibrium while
sphalerons are in equilibrium the baryon asymmetry is safe. The
out-of-equilibrium condition is
\beq
\Gamma_{\Delta L} = {\zeta(3) \over 8 \pi^3} {\lambda^4 T^3 \over M^2} <
{20 T^2 \over M_P} \simeq H
\eeq
yielding
\beq
{M \over \lambda^2} \ga 0.015 \sqrt{T_{BL} M_P}
\label{fybound}
\eeq
where $T_{BL}$ is the temperature at which the $\bl$ asymmetry was produced
or the maximum temperature when sphalerons are in equilibrium
(or the temperature $T_*$ of $e_R$ decoupling which we will momentarily
ignore) whichever
is lower.
Originally \cite{26}, $T_{BL} \sim T_c \sim 100$ GeV was chosen
giving, $ M/\lambda^2 \ga 5 \times 10^8$ GeV and corresponds
to a limit on neutrino masses $m_\nu \sim \lambda^2 v^2 / M
\la 50$ keV.  In \cite{27}, it was pointed out that sphalerons
should be in equilibrium up to $10^{12}$ GeV, in which case,
$ M/\lambda^2 \ga 10^{14}$ GeV and corresponds
to $m_\nu
\la 1$ eV. Similarly, it is possible to put constraints on
other $B$ and/or $L$ violating operators \cite{28,fish} which
include $R$-parity violating operators in supersymmetric models.
For example \cite{28}, the mass scale associated with a typical dimension
3 operator is constrained to be $m \la 2 \times 10^{-5}$ GeV,
the quartic coupling of a dimension 4 operator, $\lambda \la
7 \times 10^{-7}$ or the mass scale of a higher dimensional operator
such as a dimension 9, $\Delta B = 2$, operator is $M \ga
10^3 - 10^{13}$ GeV.  Only the latter is dependent on the choice
of $T_{BL}$.

In supersymmetric models however, it has been argued
by \cite{29}
that due to additional anomalies which can temporarily
protect the asymmetry (until the effects of supersymmetry
breaking kick in), the maximum temperature should be at $\sim 10^8 GeV$
rather than $\sim 10^{12} GeV$. Interestingly, in the context of
inflation, though the reheat temperature is typically
$10^8$ GeV, equilibration is not achieved until about $10^5$ GeV \cite{eeno}
thus the maximum
temperature should not surpass this equilibration temperature  \cite{cdo}.
These changes in $T_{max}$ would soften the limits on the mass
scales of dimension
$D \ge 5$ operators.  For example, for the D=5 ($\Delta L = 2$)
operator above,
 $M/\lambda^2 ~\ga~ 10^{9} - 10^{10}GeV$.

There are other subtleties regarding these limits.
The presence of separate lepton
asymmetries combined with mass effects can protect an asymmetry
as an equilibrium
solution\cite{krs2,dr,dko}. The rates for some operators may be
 small enough to
leave approximately conserved quantities such as
$\frac{1}{3}B - L_i$\cite{bn}.
Or, it may be possible that the asymmetry can be stored
in a weakly interacting
field such as the right-handed electron\cite{cdeo3}.

Indeed, it has been shown\cite{clkao3} that because
$e_R$ only comes into
equilibrium at the relatively cool temperature $T_*
\sim $few TeV, above $T_*$
 the baryon number is safe and the picture of baryon number
erasure is changed.  Sphaleron erasure of the
baryon asymmetry can only
occur between the $T_*$ and the decoupling temperature
$T_f$ of the additional
$B$ and/or $L$ violating rates as seen in Figure 14.
If $T_f > T_*$, the baryon asymmetry
is protected and may even be generated as shown below.
 Thus for limits on $B$ and
$L$ violating
operators, $T_{max}$ should be set at $T_*$ further relaxing
the constraints on new operators.

\begin{figure}
\vspace{8cm}
\caption{ Lepton-violating and left-right equilibrating rates.}
\end{figure}

How then can we generate a baryon asymmetry from a prior lepton asymmetry?
In addition to the mechanism described earlier utilizing a right-handed
neutrino decay,
several others are now also available.  In a supersymmetric
extension of the standard
model including a right-handed neutrino, there are numerous
possibilities. Along the
lines of the right-handed neutrino decay, the scalar partner \cite{cdo} or a
condensate \cite{mur} of ${\tilde \nu^c}$'s will easily
generate a lepton asymmetry.
Furthermore if the superpotential contains terms such
 as ${\nu^c}^3 + \nu^c H_1H_2$,
there will be a flat direction violating lepton number \cite{cdo1,cdo}
 \`{a} la Affleck and Dine.
While none of these scenarios require guts, those that
involve the out-of equilibrium
decay of either fermions, scalars or condensates must have
the mass scale of the
right-handed neutrino between $10^9$ and about $10^{11}$
GeV, to avoid washing
out the baryon asymmetry later (as can be seen from eq. (\ref{fybound}))
 and to be produced after
inflation respectively.
In contrast the decay of the flat direction condensate
(which involves other fields
in addition to ${\tilde \nu}^c$) only works for $10^{11} < M < 10^{15}$ GeV.

Flavor effects may also generate a baryon asymmetry.
Indeed consider the $\Delta L = 2$
operator discussed above. If all flavors are out of
equilibrium, then the process
for all intents and purposes can be neglected.
The only baryon asymmetry that
will result will be the small one due to mass effects \cite{krs2,dr},
unless a larger ($\ga 10^{-4}$ asymmetry is produced say by the
Affleck and Dine mechanism \cite{dko}.
  If all of the flavors are in equilibrium,
then the bound (\ref{fybound})
is not satisfied and $B$ is driven to zero. On
the other hand, if the bound is
satisfied by 1 or 2 generations, then even if initially $B = L = 0$,
 a baryon asymmetry will be generated and will be given by
\beq
B = {84 \over 247} \mu_{{2 \over 3}B - (L_1 + L_2)}
\eeq
assuming that only generations 1 and 2 are out of equilibrium and
satisfy the bound\cite{cdeo3}.

Once again, to see this more clearly it is helpful to write quantities
in terms of chemical potentials. Below $T_*$, all of the quantities of
interest can be expressed in terms of 5 chemical potentials:
$\mu_{u_L}, \mu_0,$ and $\mu_{\nu_i}$.
There are two constraints: $Q = 0$ and the sphaleron constraint
(\ref{S}) and the three initial conditions, ${1 \over 3} B - L_i$.
If all three of these conservation laws are broken eg. by the
$\Delta L = 2$ processes discussed above and $\mu_0 = - \mu_{\nu_i}$,
then we are left with two parameters, $\mu_{u_L}$ and $\mu_0$, with
two constraints: $Q = 6 \mu_{u_L} - 20 \mu_0 = 0$ and
$9 \mu_{u_L} - 3 \mu_0 =0$ yielding only the trivial solution
$B = L = \mu_{u_L} = \mu_0 = 0$.  Clearly a non-trivial solution
will be obtained when one or two of the  ${1 \over 3} B - L_i$'s
are conserved between $T_c$ and $T_*$.

Finally, a pre-existing $e_R$ asymmetry will also be
transformed into a baryon asymmetry
\cite{clkao3}.  With $e_R$ decoupled, the quantities (\ref{mus}) become
\begin{eqnarray}
B& = & 12 \mu_{u_L} \nonumber \\
L& = & 3 \mu + \mu_{e_R} - 2 \mu_H - \mu_{e_L}\\
Q& = & 6 \mu_{u_L} -2\mu - \mu_{e_R} + 13 \mu_H + \mu_{e_L} \nonumber
\end{eqnarray}
which when combined with the sphaleron condition (\ref{S}),
 and the
the $\Delta L = 2$ condition $\mu_\nu + \mu_H = 0$, one finds that
above $T_*$
\beq
B_* = {1 \over 5} \mu_{e_R} \qquad L_* = {1 \over 2} \mu_{e_R}
\eeq
independent of the initial value of $B$ and $L$.
Assuming that the $\Delta L = 2$ interactions are out of
equilibrium below $T_*$, we now have from eq. (\ref{2879}) that
\beq
B = {28 \over 79} \left( B_* - L_* \right)  \simeq -0.1 \mu_{e_R}
\eeq
Similarly, if any other other baryon and/or lepton number violating
operator was in equilibrium at some point above $T_*$ and
so long as it decouples above $T_*$, a baryon asymmetry
(or more precisely a $B - L$ asymmetry) will be produced.

In summary, I hope that it is clear that the generation
of a baryon asymmetry is
in principle relatively easy and that sphaleron interactions
may in fact aid rather than hinder the production of an asymmetry.
There are many possibilities and perhaps more than one
of them are actually responsible for the final observed asymmetry.
%
%

%
\end{document}